\documentstyle[aps,twocolumn]{revtex}

\begin{document}
\draft
\title{Effect of strain on the magneto-exciton groundstate in InP/GaInP quantum disks}
\author{K. L. Janssens\cite{karenmail}, B. Partoens\cite{bartmail} and F. M. Peeters
\cite{peetersmail}}
\address{Departement Natuurkunde, Universiteit Antwerpen (UIA), Universiteitsplein 1,
\\
B-2610 Antwerpen, Belgium}
\date{\today}
\maketitle

\begin{abstract}
The groundstate properties of an exciton in a self-assembled quantum disk are
calculated in the presence of a perpendicular magnetic field. For sufficient
wide and thin dots, the strain field leads to a confinement of the heavy hole
within the dot and the system is type I, while the light hole is confined
outside the dot and the system is type II. However, with increasing disk
thickness, the strain induces a transition of the heavy hole from inside the
disk towards the radial boundary outside the disk. For the exciton, we predict
a heavy-hole to light-hole transition as a function of the disk thickness, i.e.
forming a ``ring-like" hole wavefunction. There is a range of parameters
(radius and height of the disk) for which a magnetic field can induce such a
heavy to light hole transition. The diamagnetic shift was compared with results
from magneto-photoluminescence experiments, where we found an appreciable
discrepancy. The origin of this discrepancy was investigated by varying the
disk parameters, the valence band offset, and the effective masses.
\end{abstract}

\pacs{PACS: 73.21.La, 71.35.Ji, 85.35.Be}

\section{Introduction}
Quantum dots have been the subject of intensive studies during the last two
decades, because of their zero-dimensional nature with the corresponding
delta-function like density of states \cite{bimbergboek}. Arakawa and Sakaki
\cite{arakawa-apl40} reported in 1982 a theoretical prediction of the promising
advantages, such as e.g. a low and temperature independent threshold current,
of quantum dot lasers compared to the conventional quantum well semiconductor
laser.

The use of the Stranski-Krastanow growth mode to fabricate so-called
``self-assembled" quantum dots meant a breakthrough in the further research of
semiconductor dots. The fast and easy nature of this growth process, together
with the possibility to fabricate small and defect-free dots had a great
advantage over previous types of dots, as e.g. colloidal dots or dots created
by lithographic techniques. The formation of self-assembled dots requires two
semiconductor materials with a substantially different lattice parmeter, which
are grown on top of each other. Such lattice-mismatched hetero-epitaxy gives
rise to the formation of nanometer sized islands, governed by strain-relaxation
effects.

Confinement of charge carriers in quantum dots occurs because of the difference
in bandstructure of the two semiconductor materials. One can define two types
of quantum dots, namely type-I and type-II, depending on the band alignments.
In the case of type-I dots, both the electron and the hole are confined inside
the dot. For type-II dots, the confinement inside the dot occurs only for one
of the charge carriers, i.e. electron or hole, whereas the dot forms an
anti-dot for the other particle. Examples of type-II dots are the InP/GaInP
\cite{manus-prb62,manus-apl79} or InP/GaAs dots with the electron confined in
the dot, and the hole in the barrier, and the GaSb/GaAs \cite{muller} or
InAs/Si \cite{heitz} dots where the hole is located inside the dot, but the
electron remains in the barrier.

Up to now most research was devoted to type-I structures\cite
{polimeni,wilson,stier,brasken,bayer,ulloa,karen}, while type-II dots have
attracted fewer attention. Theoretical studies considering InP/GaInP dots were
performed by Pryor \textit{et al.} \cite{pryor} and Tadi\'{c} \textit{et al.}
\cite{milan-prb65} where a strain-dependent ${\mathbf k} \cdot {\mathbf p}$
Hamiltonian was used to calculate the electronic structure. GaSb/GaAs quantum
dots were investigated by Lelong \textit{et al.} \cite{lelong} who studied
excitons and charged excitons, and by Kalameitsev \textit{et al.}
\cite{kalameitsev} who investigated the effect of a magnetic field. In previous
work we relied on a simplified effective mass approximation, without
considering the effect of strain, to study single and coupled type-II dots in
the presence of external magnetic \cite{karen2} and electric \cite{karen3}
fields.

The formation of self-assembled quantum dots is inextricably bound up with the
occurrence of strain fields in and around the dots. The strain will have a
large impact on the bandstructure, and hereby also on the optical properties of
the dots. The hydrostatic component of the strain will shift the conduction and
valence band edges, while the biaxial strain modifies the valence bands by
splitting the degeneracy of the light- and heavy-hole bands.

Different theoretical models exist to obtain the strain distribution in and
around self-assembled quantum dots. Well-known and extensively used are the
continuum mechanical (CM) model \cite{stier,milan-prb65} and the valence force
field (VFF) model \cite{stier,williamson-prb62,milan-jap}. In the CM model the
elastic energy is minimized to obtain the distribution of the displacement in
the structure and the corresponding strain fields, whereas the VFF model is an
atomistic approach, which uses phenomenological expressions for the elastic
energy, depending on the atomic coordinates. Pryor \textit{et al.} compared the
two methods and found agreement for small strains, whereas for larger strains
discrepancies were reported \cite{pryor-jap83}. This was corroborated very
recently by Tadi\'{c} \textit{et al.}, who also compared both methods for
cylindrical InAs/GaAs and InP/InGaP dots, and found a better agreement for
InP/GaInP than for InAs/GaAs dots, where the latter has the largest lattice
mismatch.

A more simple method to calculate the strain fields near an isotropic quantum
dot was presented by Downes \textit{et al.} \cite{downes-jap81}, who used
Eshelby's theory of inclusions \cite{eshelbypaper} to express the strain
distribution in quantum dot structures. This approach was adapted by Davies,
who showed that the elastic field can be derived from a scalar potential that
obeys a Poisson equation with the lattice mismatch as charge density
\cite{davies-jap}.

In the present paper, we follow the work of Davies \cite{davies-jap} and use
the so-called isotropic elasticity model to calculate the strain fields around
a single cylindrical quantum disk. The results for this structure have recently
been shown to be in good agreement with results obtained by the CM model
\cite{milan-prb65,milan-jap}. The strain distribution is used as an input to
calculate the modification to the band structure. We use a single band
effective mass approximation to calculate the exciton properties in InP/GaInP
quantum disks in the presence of a perpendicular magnetic field, considering
both heavy- and light-hole bands. The three-dimensional confinement is known to
cause drastic effects on the bandstructure close to the zone centre, inducing a
mixing of the valence bands. However, strain causes a splitting of the heavy-
and light-hole bands, therefore reducing the mixing and making the single band
effective mass approach a justifiable first approximation \cite{bayer-prl74}.

The eight-band $\bf{k}\cdot \bf{p}$-theory and consequently also the more
simplified two band effective mass model are most accurate in the vicinity of
the $\Gamma$-point. However, the accuracy of the eight-band $\bf{k}\cdot
\bf{p}$ model decreases for smaller dots and more elaborate methods, as
empirical pseudopotential (EP) calculations are necessary. Comparison for
InAs/GaAs quantum dots shows a reasonable agreement between eight-band
$\bf{k}\cdot \bf{p}$ and EP calculations for dot base lengths $b\geq 9nm$
\cite{stier-thesis}. The contribution of ``large-$|\bf{k}|$"-states are in this
size range found to be negligible. As the dots in the present study have a
diameter of approximately $15nm$, we therefore neglect states originating from
larger $\bf{k}$-values.

The paper is organized as follows. In Sec.~II, we explain our theoretical
model. Sec.~III discusses the strain induced type-I to type-II transition for
the heavy hole, while Sec.~IV is dedicated to the heavy hole to light hole
transition of the groundstate. In Sec.~V we compare our numerical results with
experimental results of Ref.~\cite{manus-prb62}. Our results are summarized in
Sec.~VI.

\section{Theoretical model}

The strain was calculated by adapting the method used in
Ref.~\cite{davies-jap}, and which was already described in
Ref.~\cite{milan-prb65}. The elements of the strain tensor are given by
\begin{equation}\label{strain1}
  \varepsilon_{ij}({\mathbf r})=\varepsilon_0\delta_{ij}-\frac{\varepsilon_0}{4\pi}
  \frac{1+\nu}{1-\nu}\oint_{S'}\frac{(r_i-r'_i)dS'_j}{|{\mathbf r}-{\mathbf r'}|^3},
\end{equation}
where $\varepsilon_0$ is the lattice mismatch between InP and GaInP, given by
$\varepsilon_0 = (a_{InP}-a_{GaInP})/a_{GaInP}$. Furthermore, $\nu$ denotes the
Poisson ratio, generally chosen to be 1/3 for zincblende type semiconductors,
$S'$ is the surface of the dot, and $i$ runs over $x$, $y$ and $z$.

For the calculation of the exciton properties, we used a mean-field type of
approach, which is equivalent, for this problem, to the Hartree-Fock
approximation. This approach was previously introduced by us for type-II
quantum dots where strain \cite{karen2,karen1} was neglected. The coupled
single particle equations are given by
\begin{eqnarray}
\left[ H_{e(h)}-\frac{e^2}{4\pi\epsilon}\int
\frac{\rho_{h(e)}(r',z')}{|{\mathbf r}-{\mathbf r}'|}d{\mathbf r}' \right]
\psi_{e(h)}(r_{e(h)},z_{e(h)}) &&  \nonumber \\ =\epsilon_{e(h)}
\psi_{e(h)}(r_{e(h)},z_{e(h)}), &&
\end{eqnarray}
with $H_e$ ($H_h$) the single particle Hamiltonian for the electron (hole). In
the presence of strain, the single particle Hamiltonian for the electron can be
written as
\begin{eqnarray}
  H_e&=&-\frac{\hbar^2}{2}\nabla
  \frac{1}{m^*({\mathbf r})}\nabla
  +V_{c}({\mathbf r})+a_{c}\varepsilon_{hy}({\mathbf r}) \nonumber
  \\
  &&+\frac{\hbar^2}{2m^*({\mathbf r})}\frac{l_{e}^2}{r^2}+\frac{l_e}{2}\hbar\omega_{c,e}
  +\frac{1}{8}m^*({\mathbf r})\omega_{c,e}^2r^2,
\end{eqnarray}
with $\omega_{c,e}({\mathbf r})=eB/m^*({\mathbf r})$, where $m^*({\mathbf r})$
is the position dependent electron effective mass, $a_c$ is the conduction band
deformation potential, $l_e$ is the electron angular momentum, and the
hydrostatic strain, $\varepsilon_{hy}$, is defined by
$\varepsilon_{xx}+\varepsilon_{yy}+\varepsilon_{zz}$.

Furthermore, we make a distinction between the heavy and the light hole, as the
strain will act differently on them. For the heavy hole, we have
\begin{eqnarray}
  H_{hh}({\mathbf r})&=&\frac{\hbar^2}{2}\left\{\nabla_{\|}\frac{1}{m_{hh,\|}({\mathbf r})}
  \nabla_{\|}
  +\nabla_{z}\frac{1}{m_{hh,z}({\mathbf r})}\nabla_{z}\right\}
  \nonumber
  \\ &&+V_{h}({\mathbf r})+a_{v}\varepsilon_{hy}({\mathbf r})+b
  ((\varepsilon_{xx}+\varepsilon_{yy})/2+\varepsilon_{zz}) \nonumber
  \\ &&+\frac{\hbar^2}{2m_{hh,\|}({\mathbf r})}\frac{l_{h}^2}{r^2}-\frac{l_h}{2}\hbar\omega_{c,hh}
  +\frac{1}{8}m_{hh,\|}({\mathbf r})\omega_{c,hh}^2r^2,
\end{eqnarray}
with $\omega_{c,hh}({\mathbf r})=eB/m_{hh,\|}({\mathbf r})$, and
$m^{-1}_{hh,\|}({\mathbf r})=[\gamma_{1}({\mathbf r})+\gamma_{2}({\mathbf r})]$
the in-plane heavy hole effective mass, $m^{-1}_{hh,z}({\mathbf
r})=[\gamma_{1}({\mathbf r})-2\gamma_{2}({\mathbf r})]$ the heavy hole
effective mass along the $z$-direction. The valence band deformation potentials
are given by $a_v$ and $b$, and $\gamma_{i}({\mathbf r})$ are the Luttinger
parameters.

For the light hole, this becomes
\begin{eqnarray}
  H_{lh}({\mathbf r})&=&\frac{\hbar^2}{2m_{0}}\left\{\nabla_{\|}
  \frac{1}{m_{lh,\|}({\mathbf r})}\nabla_{\|}
  +\nabla_{z}\frac{1}{m_{lh,z}({\mathbf r})}\nabla_{z}\right\}
  \nonumber
  \\ &&+V_{h}({\mathbf r})+a_{v}\varepsilon_{hy}({\mathbf r})-b
  ((\varepsilon_{xx}+\varepsilon_{yy})/2+\varepsilon_{zz}) \nonumber
  \\ &&+\frac{\hbar^2}{2m_{lh,\|}({\mathbf r})}\frac{l_{h}^2}{r^2}-\frac{l_h}{2}\hbar\omega_{c,lh}
  +\frac{1}{8}m_{lh,\|}({\mathbf r})\omega_{c,lh}^2r^2.
\end{eqnarray}
with $\omega_{c,lh}({\mathbf r})=eB/m_{lh,\|}({\mathbf r})$, and
$m^{-1}_{lh,\|}({\mathbf r})=[\gamma_{1}({\mathbf r})-\gamma_{2}({\mathbf r})]$
the in-plane light hole effective mass, and $m^{-1}_{lh,z}({\mathbf
r})=[\gamma_{1}({\mathbf r})+2\gamma_{2}({\mathbf r})]$ the light hole
effective mass along the $z$-direction.

We solved the coupled single-particle equations self-consistently using an
iterative procedure. More information about the implementation and numerical
procedure can be found in Ref.~\cite{karen1}. After convergence of the
iteration procedure, the total energy is given by
\begin{equation}
E_{exciton}=E_{e}+E_{h}+\frac{e^{2}}{4\pi \epsilon } \int \int \frac{\rho
_{e}(r,z)\rho _{h}(r^{\prime },z^{\prime })}{|{\mathbf r}- {\mathbf r}^{\prime
}|}d{\mathbf r}d{\mathbf r}^{\prime },
\end{equation}
where $E_e$ and $E_h$ are the single particle energies.

For our numerical calculation, we took the material parameters of InP/GaInP, as
used in Refs.~\cite{pryor,milan-prb65}, which are given in Table~I. Here
$m_{hh,\|}$, $m_{hh,z}$, $m_{lh,\|}$, and $m_{lh,z}$ are calculated from the
Luttinger parameters using the formulas mentioned above. From Table~I follows
that the valence band offset for this system is \textit{negative}
($V_h=-45meV$), i.e. we have a type-II system. This appeared to be rather
controversial, however, as a theoretical calculation based on first principles
by Wei and Zunger \cite{zungerwei-apl72} showed that the band offset should be
positive ($V_h=55meV$), leading to a type-I confinement. Experimentally, Hayne
\textit{et al.} \cite{manus-prb62,manus-apl79} conclude that the hole should be
located in the barrier, and that the system is type-II. In a first step, we
will use the negative band offset, as we used previously. Later, we will
investigate how our results are modified when we take the positive bulk hole
band offset.

\section{Strain induced type-I - type-II transition for the heavy hole.}

As follows from the equations, hydrostatic strain shifts the conduction and
valence band offsets. Furthermore, the biaxial strain induces a splitting of
the heavy- and light-hole bands. The heavy-hole band is shifted towards higher
energies, whereas the light-hole band shifts towards lower energies.

The unstrained valence band offset is slightly negative, i.e. $V_h=-45meV$, but
the heavy-hole band offset will become strongly positive after strain is
introduced. The heavy hole will thus be strongly confined inside the disk,
making the system type-I. In contrast, the light-hole band decreases in energy,
making the system even more type-II and the light hole will be located outside
the disk. The numerical results for a disk of radius $R=8nm$ and thickness
$d=4nm$ are shown in Fig.~1. The unstrained valence band offset (solid curve)
is depicted together with the strained heavy hole (dashed curve) and light hole
(dotted curve) band offsets, in the direction perpendicular to the disk along
its center (Fig.~1(a)) and along its radial (Fig.~1(b)). Contourplots of the
heavy and light hole potentials are depicted as insets in Fig.~1. For the heavy
hole, we find a deep potential maximum in the middle of the quantum disk.
Notice that in the radial direction, this potential decreases towards the disk
boundary and increases again just after the radial disk boundary and can have
even a higher maximum, but which in the considered case is narrower. The light
hole potential shows the deepest confinement (i.e. highest potential) in the
barrier material, just on top and below the disk, which will be the
preferential position for the light hole.

We found that when we increased the thickness $d$ of the disk, the heavy hole
moves towards the radial boundary of the disk. This effect is purely due to the
strain, which increases the potential maximum at the radial boundary with
respect to the potential inside the disk (see Fig.~2(b)). For thicker disks,
the height of the potential \textit{in} the disk systematically decreases,
making it eventually preferable for the heavy hole to move out of the disk and
the system becomes type-II also for the heavy hole. Fig.~2(a) depicts the
downward shift of the confinement potential for increased $d$ along the
$z$-direction. We summarized our results in the phase diagram in Fig.~3. To
construct this figure, we calculated the probability of the heavy hole to sit
at the radial boundary of the disk
\begin{equation}
P_{side}=2\pi \int_{-\infty }^{\infty }dz_{h}\int_{R}^{\infty
}dr_{h}~r_{h}\left| \Psi _{hh}(r_{h},z_{h})\right| ^{2}.
\end{equation}
The line in Fig.~3 indicates the probability $P_{side}=50\%$. At the right side
of this line, the heavy hole is mainly located inside the disk (Fig.~3(a)), and
the system is type-I-like. At the left side of the fifty percent line, the
heavy hole is predominantly located at the radial boundary outside of the disk
(Fig.~3(b)), and the system is type-II-like.

From our previous work \cite{karen2,karen1}, we know that when the hole is
located at the radial boundary of the disk it will lead to the occurrence of
hole angular momentum transitions as a function of a magnetic field applied
parallel to the growth ($z$) direction. Such angular momentum transitions also
occur for the exciton and are not found for type-I excitons.

\section{Heavy hole to light hole transition of the groundstate.}

In the present approach, mixing of the valence bands is neglected and we thus
solve two separate sets of coupled equations, namely one for the heavy hole and
one for the light hole. Next we will compare the energy of the light and heavy
hole exciton in order to determine which one is the ground state. The results
are summarized in Fig.~4, which shows a phase diagram of the heavy to light
hole transition as a function of both $R$ and $d$. We find that the light hole
exciton becomes the groundstate for increasing thickness of the disk. The full
curve in Fig.~4 shows the separation between the heavy and light hole exciton
groundstate in the absence of a magnetic field. The dotted curve is the result
of Fig.~3, which separates the heavy hole type-I and type-II groundstate.
Notice that the heavy hole exciton is in the heavy hole type-I region for
$R>6.4nm$. When the heavy hole is at the radial boundary outside the dot, the
exciton energy is higher than for the (type-II) light hole exciton, except in a
very narrow region of $d$-values for $R<6.4nm$.

The dashed curve shows the result for $B=50T$. Thus a magnetic field lowers the
light hole groundstate with respect to the heavy hole. This is made more clear
in Fig.~5, where the evolution of the heavy (full curve) and light hole (dashed
curve) exciton energies are shown as function of magnetic field for a fixed
disk size, i.e. $R=6.5nm$ and $d=3.5nm$. The reason for this \textit{magnetic
field induced heavy-to-light hole transition} can be found in the stronger
effect of the magnetic field on the heavy hole. This is due to a larger
magnetic energy $\hbar\omega_{c,hh}=\hbar eB/m_{hh,\|}$, which is caused by a
smaller in-plane heavy hole mass, known as \textit{mass-reversal}
\cite{cardonabook}. The insets show respectively the heavy hole wavefunction at
$B=0T$ (a) and the light hole wavefunction at $B=50T$ (b). Thus this
heavy-light hole transition is accompanied with a spatial direct-to-indirect
exciton transition.

It is also possible to have a magnetic field induced heavy-to-light hole
transition where both heavy and light hole exciton states are spatially
indirect. This is shown in Fig.~6, where the evolution of the heavy (full
curve) and light (dashed curve) hole exciton are depicted as function of
magnetic field for a disk with radius $R=5.5nm$ and thickness $d=3.5nm$. In the
insets are depicted both the probability density $|\Psi|^2$ (a,b,e) and the
radial position probability density $r|\Psi|^2$ (c,d,f) for the hole in the
exciton groundstate, for different values of the magnetic field, as indicated.
The heavy hole is shown in Figs.~6(a-d), where from the radial position
probability density follows that indeed the system is type-II. An increasing
magnetic field pushes the heavy hole wavefunction more inside the disk, until
it would eventually become type-I. However, before this happens, the light hole
exciton becomes the groundstate of the system, which is still type-II, as shown
in Figs.~6(e,f).

\section{Comparison with experiment.}

In order to compare our theoretical results with the magneto-photoluminescence
experiments of \cite{manus-prb62}, we have to calculate the exciton transition
energy as a function of the magnetic field. The transition energy is defined as
follows:
\begin{equation}\label{extrans}
  E_{trans} = E_{exciton}+E_g
\end{equation}
where $E_g$ is the bandgap energy of the disk material and $E_{exciton}$
(Eq.~6) contains the single particle electron and hole energies and the Coulomb
binding energy. Our theoretical results (curves) are compared with the
experimental photoluminescence results (squares) from Hayne \textit{et al.}
\cite{manus-prb62} in the inset of Fig.~7. We show both the heavy (full curve)
and light (dashed curve) hole exciton energy, for two values of the disk
thickness: $d=2nm$ and $d=2.5nm$, as indicated on the plot (the disk radius was
taken $R=8nm$ for all curves). A change in the thickness almost uniformly
shifts the curves up or down in energy, which is due to a corresponding change
of the confinement energy (mostly of the electron). For these sets of disk
parameters, the heavy hole exciton is the ground state. But we find that the
light hole exciton for $d=2.5nm$ approaches most closely the experimental
result for small $B$, while the heavy hole results are closer for high $B$. By
fine tuning the value of $d$, it is possible to fit the $B=0T$ heavy hole
transition energy to the experimental result. To show more clearly the magnetic
field dependence, we investigated the exciton diamagnetic shift, defined as
\begin{equation}\label{exdia}
  \Delta E=E_{trans}(B)-E_{trans}(B=0T).
\end{equation}
The result is depicted in Fig.~7 for $d=2.5nm$ by the thick full (heavy hole)
and dashed (light hole) curves. Notice that the diamagnetic shift of the
exciton is overestimated by more than a factor two in the considered magnetic
field range.

In order to explain the discrepancy, we systematically investigated the
influence of different disk and material parameters on the diamagnetic shift of
the exciton. The results for $d=2nm$ are on top of the $d=2.5nm$ results. This
is not too surprising because the disk thickness only minorly influences the
in-plane motion. But a decrease of the disk radius will decrease the
diamagnetic shift. However, even a substantial decrease to $R=6nm$ can not
explain the experimental results quantitatively, as shown in Fig.~7 by the thin
curves, although the theoretical results are closer to the experimental results
than for $R=8nm$. We find that the light hole exciton is most strongly affected
and shows a smaller diamagnetic shift. However, the shift is still large for
high fields, and can thus not be the main reason for the discrepancy. But from
the phase diagram (Fig.~4) we know that the heavy hole exciton is still the
groundstate and its diamagnetic shift decreased only by 7.7\%. It would not be
realistic to further decrease the disk radius, as measurements found a disk
diameter of approximately $16nm$.

Next, the influence of the masses was investigated. In order to obtain a
smaller diamagnetic shift, an increase of the masses is needed: the higher the
mass, the stronger the particles are bound in the disk, and the smaller will be
the influence of the magnetic field, i.e. the diamagnetic shift is inversely
proportional to the exciton mass. For the electron, we increased the mass with
a factor of two, namely $m^*_{e} (InP) = 0.15m_0$. As the magnetic field acts
mainly on the radial direction, a change of the in-plane mass ($m_{\|}$) will
have the largest effect on the diamagnetic shift. We will therefore only vary
this in-plane mass for the holes. Furthermore, we found that the heavy hole is
located in the disk, whereas the light hole is located in the barrier material.
For the heavy hole exciton we thus vary only the in-plane mass \textit{in the
disk}, while for the light hole exciton we vary only the in-plane mass
\textit{in the barrier}. The results are shown by the dotted curve in Fig.~7.
An excellent fit to the experimental diamagnetic shift with the heavy hole
exciton could be found (disk radius $R=8nm$ and thickness $d=2.5nm$) for
$m_{hh,\|}=0.5m_0$, whereas for the light hole, we found a good fit for
$m_{lh,\|}=1.50m_0$.

The assumption of a negative band offset was questioned in
Ref.~\cite{zungerwei-apl72} where a positive unstrained valence band offset of
$55meV$ was found theoretically. The result for the transition energy with a
positive unstrained hole band offset of $V_h=55meV$ is shown in the inset of
Fig.~8, where we find that the heavy hole now more closely approximates the
experimental result, although the quantitative agreement is less good than for
a negative unstrained offset. Notice also the large difference between the
results for heavy and light hole. The reason is that the heavy hole is now much
stronger confined inside the disk. The corresponding diamagnetic shift is shown
in Fig.~8 for $R=8nm$ and $d=2.5nm$, where the results for the negative offset
are shown by the thick curves, whereas the results for the positive offset are
shown by the thin curves. We notice that the heavy hole is stronger affected by
the reversed offset: the diamagnetic shift at $B=50T$ has decreased by $\sim
5meV$. We attribute this to the stronger confinement of the heavy hole inside
the disk: as the hole will already be confined in the unstrained case, strain
will only enhance this effect for the heavy hole. Furthermore a stronger
potential confinement decreases the effect of the magnetic field on the
particle. The light hole, on the contrary, is only slightly effected by the
altered band offset. Although the height of the potential barrier for the light
hole has decreased, it is still appreciable, and will not have a drastic
influence on the magnetic field behaviour of the light hole. However, the
qualitative discrepancy still exists in Fig.~8, which can thus not be explained
by the reversed band offset. Also for the case of the positive unstrained VB
offset, we investigated the influence of the masses. Again we took a fixed
electron mass in the dot of $m^*_e=0.15m_0$, both for the heavy hole exciton as
for the light hole exciton. Furthermore, following the same reasoning as above,
we only fitted the in-plane heavy and light hole masses. We found an excellent
fit to the experimental results for $m_{hh,\|}=0.5m_0$ and $m_{lh,\|}=3.0m_0$
for respectively the heavy and the light hole exciton, which is shown in Fig.~8
by the dotted curve. Notice that in this case we need a light hole effective
mass which is two times larger than in the case of a negative unstrained
offset. We attribute this to the fact that the light hole is more weakly bound,
and that thus a higher mass is needed to obtain a smaller diamagnetic shift.

From the above results, we can conclude that a change of the disk parameters
(i.e. radius and/or thickness) is not the main origin of the discrepancy
between theory and experiment. Furthermore, we find that taking a positive
unstrained valence band offset has only a minor influence to our theoretical
results. However, both for the positive and for the negative unstrained offset,
a change of the electron mass and of the in-plane hole masses could change
drastically the diamagnetic shift, and a good fit to the experimental result
could be obtained. Effects due to disorder may have an appreciable effect on
the masses and are expected to increase them. But on the other hand the use of
the effective mass approximation is questionable in such a situation.

\section{Conclusions}
We have investigated theoretically the properties of an exciton in a strained
self-assembled quantum dot. The dot was modelled by a disk and strain fields
were taken into account using the isotropic elasticity model. The exciton
energy and wavefunction were obtained by means of a mean field theory using the
Hartree approximation.

Investigation of the effect of strain on the confinement potentials learns
that, starting from a negative (type II) unstrained valence band offset, the
confinement potential for the heavy hole is reversed (i.e. becomes type I),
while for the light hole the system becomes even stronger type II. However,
with increasing disk thickness, the heavy hole potential inside the disk was
found to decrease strongly with respect to a potential maximum at the radial
boundary outside the disk. For a certain disk thickness, it becomes thus
preferential for the heavy hole to move towards the radial boundary, into the
barrier material, making the system also type II. A consequence of this
behaviour is the occurrence of angular momentum transitions for the heavy hole
when a magnetic field is applied.

A phase diagram was constructed as a function of the disk radius and thickness,
indicating whether the heavy hole exciton or the light hole exciton forms the
groundstate of the system. This was done both for $B=0T$ and $B=50T$. The
magnetic field was found to be able to induce a heavy-to-light hole transition.
This magnetic field induced transition was attributed to the heavy hole mass,
which has a smaller in-plane mass than the corresponding light-hole mass.

Finally, both the transition energy and the energy of the diamagnetic shift
were calculated as a function of the magnetic field, and compared with
experimental results by Hayne \textit{et al.} \cite{manus-prb62,manus-apl79}.
An appreciable discrepancy was found for the diamagnetic shift, and the origin
of this discrepancy was investigated. We considered a smaller disk radius and
found that this slightly decreases the difference between theory and
experiment. Next, the masses were strongly increased, and we found that we can
fit both the heavy and the light hole exciton energies to the experimental
result. Furthermore, instead of taking a negative unstrained valence band
offset, we investigated the effect of starting with a positive unstrained
offset. We found that this gives only a minor change in the theoretical
results, and that again an increase of the masses is needed to obtain a good
fit.

As far as the experimental result is concerned, we may conclude that from the
diamagnetic shift, we find that the most plausible explanation for the small
shift is an increase of the electron mass by a factor of two, and as a
consequence its value is closer to the GaInP mass, and an increase of the
in-plane heavy hole mass such that its value is close to its perpendicular
mass. Such increases in electron and hole masses are not unrealistic for a
disordered quantum dot system. Thus the exciton observed by Hayne \textit{et
al.} \cite{manus-prb62} is most probably a type-I heavy hole exciton in a
strongly disordered quantum dot. In order to explain the experimental results
in terms of a light hole exciton we have to increase the light hole in-plane
mass with approximately a factor of ten, which is rather unrealistic.
Furthermore, it is the heavy hole exciton which is the ground state of the
system for the experimental value of the sizes of the quantum dot.

\section{Acknowledgements}
K. L. J. is supported by the ``Instituut voor de aanmoediging van Innovatie
door Wetenschap en Technologie in Vlaanderen'' (IWT-Vl) and B. P. is a
post-doctoral researcher with the Flemish Science Foundation (FWO-Vl). Part of
this work was supported by the FWO-Vl, The Belgian Interuniversity Attraction
Poles (IUAP), the Flemish Concerted Action (GOA) Programme, the University of
Antwerp (VIS) and the European Commission GROWTH programme NANOMAT project,
contract No. G5RD-CT-2001-00545. We acknowledge interesting discussions with M.
Tadi\'{c}, M. Hayne, A. Zunger, A. Matulis and J. Davies.

\begin{table}
\caption{Material parameters, taken from Ref.~[15].}
\begin{tabular}{lcc}
  Parameter & InP & Ga$_{0.51}$In$_{0.49}$P \\
  \hline
  $m^*_{e} (m_0)$ & 0.077 & 0.125 \\
  $\gamma_1$ & 4.95 & 5.24 \\
  $\gamma_2$ & 1.65 & 1.53 \\
  $m_{hh,\|}$ & 0.1515 & 0.1477 \\
  $m_{hh,z}$ & 0.606 & 0.4587 \\
  $m_{lh,\|}$ & 0.303 & 0.269 \\
  $m_{lh,z}$ & 0.121 & 0.1205 \\
  $E_g (eV)$ & 1.424 & 1.97 \\
  $V_c (eV)$ & 1.379 & 1.97 \\
  $V_v (eV)$ & -0.045 & 0.0 \\
  $a_c (eV)$ & -7.0 &  \\
  $a_v (eV)$ & 0.4 &  \\
  $b (eV)$ & -2.0 &  \\
  $\epsilon$ & 12.61 & 12.61 \\
  $a (nm)$ & 0.58687 & 0.56532 \\
\end{tabular}
\end{table}

\begin{figure}[tbp]
\caption{The confinement potentials for the hole along the $z$-direction (a)
and along the radial direction (b) for a disk with radius $R=8nm$ and thickness
$d=4nm$. The unstrained valence band offset is shown by the solid curve,
whereas the dashed and dotted curves denote respectively the heavy and light
hole confinement potentials. The insets show contourplots of the heavy (left
panel) and light (right panel) hole confinement potentials. The white regions
denote the highest potential, where the respective hole will be located.}
\end{figure}

\begin{figure}[tbp]
\caption{Heavy hole confinement potentials along the radial (a) and $z$ (b)
direction, for a disk with radius $R=8nm$ and thickness $d=2nm$ (solid curve)
and $d=8nm$ (dashed curve).}
\end{figure}

\begin{figure}[tbp]
\caption{Phase diagram for the probability for the heavy hole to be located at
the radial boundary of the disk, as a function of the disk radius and
thickness. The curve denotes a probability of 50\%. The insets show the heavy
hole wavefunctions for respectively a type I-like system (a) and a type II-like
system (b).}
\end{figure}

\begin{figure}[tbp]
\caption{Phase diagram indicating whether the heavy or light hole exciton is
the groundstate, as a function of the disk radius and thickness. The solid
curve denotes the result for $B=0$, whereas the dashed curve gives the result
for $B=50T$. The dotted line indicates the 50\% probability for the heavy hole
to be located at the radial boundary of the dot for $B=0$.}
\end{figure}

\begin{figure}[tbp]
\caption{The exciton energy as a function of the magnetic field for a disk with
radius $R=6.5nm$ and thickness $d=3.5nm$, both for the heavy hole exciton
(solid curve) and the light hole exciton (dashed curve.) The insets show
contourplots of the probability densities of respectively the heavy hole at
$B=0T$ (a) and the light hole at $B=50T$ (b).}
\end{figure}

\begin{figure}[tbp]
\caption{The exciton energy as a function of the magnetic field for a disk with
radius $R=5.5nm$ and thickness $d=3.5nm$, both for the heavy hole exciton
(solid curve) and the light hole exciton (dashed curve.) The insets show
contourplots of respectively the probability densities and radial probability
densities of the heavy hole at $B=0T$ (a,c) and at $B=10T$ (b,d), and of the
light hole at $B=20T$ (e,f).}
\end{figure}

\begin{figure}[tbp]
\caption{The exciton diamagnetic shift as a function of the magnetic field, for
a disk with thickness $d=2.5nm$ and for respectively $R=8nm$ (thick curves) and
$R=6nm$ (thin curves). The solid curves indicate the heavy hole exciton,
whereas the dashed curves indicate the light hole exciton. The dotted curve
depicts the result with fitted masses. The squares denote the experimental
result of Ref.~[3]. The inset depicts the exciton transition energy for a disk
with radius $R=8nm$ and thickness $d=2nm$ and $d=2.5nm$, as indicated in the
plot.The solid curve denotes the heavy hole exciton, whereas the dashed curve
gives the result for the light hole exciton.}
\end{figure}

\begin{figure}[tbp]
\caption{The exciton diamagnetic shift as a function of the magnetic field for
both a positive (thick curves) and negative (thin curves) unstrained valence
band offset. The results for the heavy and light hole exciton are given by
respectively the solid and dashed curves, while the squares indicate the
experimental result of Ref.~[3]. The dotted curve denotes the results with
fitted masses. Inset: The exciton transition energy as function of the magnetic
field. The same convention for the curves is used as in the main figure.}
\end{figure}

\end{document}